\documentclass[showpacs,preprintnumbers,amsmath,amssymb]{revtex4}
\usepackage{graphicx}
\usepackage{bm}
\usepackage{amsfonts}
\usepackage{indentfirst}
\usepackage[small]{caption2}
\usepackage{longtable}

\begin{document}


\title{Dissipative dynamics of nondegenerate two-photon Jaynes-Cummings model}
\author{E.K. Bashkirov}
 \altaffiliation[Electronic adress:]{bash@ssu.samara.ru}
\author{M.S.Rusakova}%
 \email{ruma@ssu.samara.ru}
\affiliation{%
Department of General and Theoretical Physics, Samara State University, Acad. Pavlov Str.1 , 443011 Samara, Russia
\\
}%

\date{\today}

\begin{abstract}
A nondegenerate two-photon Jaynes-Cummings model is investigated where the leakage of photon through the cavity is
 taken into account. The effect of cavity damping on the mean photon number,
 atomic populations, field statistics  and both field and atomic squeezing
 is  considered  on the basis of master equation
 in dressed-state approximation  for initial coherent  fields and excited atom.
\end{abstract}

\pacs{42.50.Ct, 42.50.Dv, 42.50.Hz }
\maketitle

\section{Introduction}
Over the last two decades much attention has been focused on the properties of the dissipative variants
 of the Jaynes-Cummings model (JCM). The theoretical efforts has been stimulated by experimental progress
 in investigation of the interaction of a single atom with electromagnetic field inside the cavity \cite{1}.
 The experiments with highly excited Rydberg atoms allowed some of the predictions of the extended version
 of JCM to be proved. Besides the experimental drive, there also exists  a theoretical motivation to include
 relevant damping mechanism to JCM because its dynamics becomes more interesting. The dissipative effects in
 JCM caused by the energy exchange between the system and environment have been studied both
 analytically \cite{2}-\cite{7} and numerically \cite{8}-\cite{10}. Last few years the JCM with phase
 damping, as applied to decoherence and entanglement, has been also treated intensively \cite{11,12}.

It's known that two-photon processes are very important in atomic systems due to high degree of correlation
 between the emitted photons. Hence, one valuable extension of the JCM is well-known two-photon JCM. With
 the experimental realization of two-photon micromaser \cite{13} the dissipative two-photon JCM has attracted a
  great deal of attention \cite{14}-\cite{18}. It is worth nothing that in all mentioned works the dissipative
  dynamics of the degenerate two-photon JCM has been under consideration. In order to advance one step further
  in the investigation of two-photon processes, the JCM with two-mode two-photon interaction or
  nondegenerate two-photon JCM (NTPJCM) were proposed.  A remarkable feature of such a model is that one mode
  can be used to affect on the other mode. The lossless NTPJCM has been used to study time evolution of
  the atomic and photon operators, the second-order coherence function, the one- and two-mode squeezing,
  the atomic dipole squeezing, the emission spectra and quantum entropy and entanglement
  without and with consideration of Stark-shifts \cite{19}-\cite{31}. The influence of phase damping
  on nonclassical properties of NTPJCM has been considered in \cite{17}. The effect of a cavity damping on the time behaviour of the atomic
  population in  the special case when the fields are initially in the two-mode squeezed vacuum has been taken investigated
   by Gou  \cite{19}.
It's of great interest to
  investigate the role of energy dissipation in dynamics of the NTPJCM for arbitrary fields states.

\section{Model Hamiltonian and kinetic equations }

The nondegenerate two-photon Jaynes-Cummings model is an effective
 two-level atom with upper and lover states denoted
 $|e\rangle$ and $|g\rangle$ respectively interacting with two modes of quantum electromagnetic
 field with frequencies $\omega_1$ and $\omega_2$ through two-photon transition. The Hamiltonian for
 such a system in dipole and RWA approximation is

\begin{equation}
H = \hbar \omega_0 R^z + \hbar (\omega_1 a_1^+ a_1+\omega_2 a_2^+ a_2)+ \hbar g (a_1 a_2 R^+ + a_1^+ a_2^+R^-),
\end{equation}
where $\omega_0$ is the atomic transition frequency, $\omega_1$ and $\omega_2$ are the cavity mode frequencies, $a_i^+$ $(a_i)$ are the creation (annihilation) operator of the photon ($i = 1,2$), $R^z$ is the inversion population operator,
 $R^{\pm}$   are the operators describing the transitions between the upper and lower levels and $g$ is the atom-field
 coupling constant. We have ignored the Stark shift caused by the intermediate level and denoted the detuning parameter as
 $$\Delta = \omega_0-\omega_1 - \omega_2 .$$
which satisfy the condition $\Delta \ll \omega_0, \omega_1, \omega_2 $.

In order to describe dissipation one has to treat the system as open. In this paper we take into account only the field
 mode damping and ignore the atom damping. The cavity is assumed to be at zero temperature. Then, the master equation
  for the density matrix of combined (atom-field) system is

\begin{equation}
\frac{\partial \rho}{\partial t} = - i/\hbar\,[H, \rho]
 - \sum_{i=1}^2 k_i\left (a_i^+ a_i\rho - 2\,a_i\rho\,a_i^+ + \rho\,a_i^+ a_i \right ),
\end{equation}
where $2 k_i\quad (i=1,2)$ are the rates of photon leakage  from the cavity. For the sake of simplicity
 we put $k_1=k_2=k$.

Using the representation
 $$W(t)=e^{\frac{i}{\hbar}Ht}\rho (t)e^{-\frac{i}{\hbar}Ht},\qquad O(t)=e^{\frac{i}{\hbar}Ht} O e^{-\frac{i}{\hbar}Ht},$$
where $O$ is an arbitrary operator of combined system, one can rewrite the master equation (2) in the form
\begin{equation}
\frac{\partial W}{\partial t} = -
 \sum_{i=1}^2 k\left (a_i^+ a_iW - 2\,a_iW\,a_i^+ + W\,a_i^+ a_i \right ),
\end{equation}

To solve Eq.(3) we have used the so-called dressed-states representation, i.e. representation consisting of the complete set of hamiltonian eigenstates. For lossless cavity the full set of dressed states are
\begin{equation}
 |\Psi_{n}^{\pm}\rangle =
\frac{\gamma^\pm_{n_1n_2}}{\sqrt{2}}|+,n_1,n_2\rangle\pm \frac{\gamma^\mp_{n_1n_2}}{\sqrt{2}}|-,n_1+1,n_2+1\rangle,
\end{equation}
with eigenvalues
$$E_{n}^{\pm} =\hbar\phi_{n_1n_2}\pm\hbar\Omega_{n_1n_2},$$
where $$\phi_{n_1n_2}=\omega_1(n_1+\frac12)+\omega_2(n_2+\frac12),$$
$$\Omega_{n_1n_2}=\sqrt{\frac{\Delta^2}{4}+g^2(n_1+1)(n_2+1)}, \quad\delta(n)=\Delta/\Omega(n).$$
Here $|\alpha; n>$ refers to a state with $n$ photons in the cavity field mode and the atom in the excited ($\alpha=+$) or in the ground ($\alpha=-$) state
 $$|\alpha; n> = |\alpha>_A\,|n>_{F}, $$
 where  $n= 0, 1, 2, \dots$.

For finite-Q cavity the above states (4) should be added  with the states:  $ |\Psi^{l}_1\rangle = |-,1,0\rangle$, \quad $E=\hbar\omega_1-\frac12\hbar\omega_0$,

$ |\Psi^{l}_2\rangle = |-,0,1\rangle$, \quad $E=\hbar\omega_2-\frac12\hbar\omega_0$,

$ |\Psi^{l}_3\rangle = |-,0,0\rangle$, \quad $E=-\frac12\hbar\omega_0,$ which take into account the  photon leakage with no atom change.

Using the secular approximation which holds for $2 k n^2 \ll g\sqrt{n+1}$ \cite{3}, i.e. neglecting the oscillatory terms, the equations for the diagonal elements of density matrix $W$ are found to be
\begin{equation}
\langle\Psi^\pm_{n_1n_2}|\dot W|\Psi^\pm_{n_1n_2}\rangle= -k\left\{2(n_1+n_2)+\gamma_{n_1n_2}^{\mp^2}\langle\Psi^\pm_{n_1n_2}|W|
\Psi^\pm_{n_1n_2}\rangle-\right.\phantom{2(n_1+n_2)}
\end{equation}
$$-\frac{g^2(n_2+1)}{2}\left[\frac{n_1+1}{\Omega_{n_1n_2}}
\frac{\gamma^\pm_{n_1+1,n_2}}{\gamma^\mp_{n_1n_2}}+ \frac{n_1+2}{\Omega_{n_1+1,n_2}}\frac{\gamma^\mp_{n_1n_2}}{\gamma^\pm_{n_1+1,n_2}}\right]^2 \langle
\Psi^\pm_{n_1+1,n_2}|W|\Psi^\pm_{n_1+1,n_2}\rangle-$$
$$-\frac{g^2(n_2+1)}{2}\left[\frac{n_1+1}{\Omega_{n_1n_2}}
\frac{\gamma^\mp_{n_1+1,n_2}}{\gamma^\mp_{n_1n_2}}- \frac{n_1+2}{\Omega_{n_1+1,n_2}}\frac{\gamma^\mp_{n_1n_2}}{\gamma^\mp_{n_1+1,n_2}}\right]^2 \langle \Psi^\mp_{n_1+1,n_2}|W|\Psi^\mp_{n_1+1,n_2}\rangle-$$
$$-\frac{g^2(n_1+1)}{2}\left[\frac{n_2+1}{\Omega_{n_1n_2}}
\frac{\gamma^\pm_{n_1,n_2+1}}{\gamma^\mp_{n_1n_2}}+ \frac{n_2+2}{\Omega_{n_1,n_2+1}}\frac{\gamma^\mp_{n_1n_2}}{\gamma^\pm_{n_1,n_2+1}}\right]^2 \langle \Psi^\pm_{n_1,n_2+1}|W|\Psi^\pm_{n_1,n_2+1}\rangle-$$
$$-\frac{g^2(n_1+1)}{2}\left[\frac{n_2+1}{\Omega_{n_1n_2}}
\frac{\gamma^\mp_{n_1,n_2+1}}{\gamma^\mp_{n_1n_2}}- \frac{n_2+2}{\Omega_{n_1,n_2+1}}\frac{\gamma^\mp_{n_1n_2}}{\gamma^\mp_{n_1,n_2+1}}\right]^2 \langle \Psi^\mp_{n_1,n_2+1}|W|\Psi^\mp_{n_1,n_2+1}\rangle,\phantom{-}$$ where
$$\gamma^\pm_{n_1n_2}=\sqrt{1\pm\frac{\Delta}{2\Omega_{n_1n_2}}}\phantom{l}$$
and
\begin{eqnarray}
\langle\Psi^l_i|\dot W|\Psi^l_i\rangle&=& -k(2\langle\Psi^l_i|W|\Psi^l_i\rangle-\frac{g^2}{\Omega_{00}^2
\gamma_{00}^{+^2}}\langle\Psi^+_{00}|W|\Psi^+_{00}\rangle-\\
&-&\frac{g^2}{\Omega_{00}^2\gamma_{00}^{-^2}}\langle\Psi^-_{00}|W|\Psi^-_{00}\rangle \qquad (i=1,2),\nonumber\\
\langle\Psi^l_3|\dot W|\Psi^l_3\rangle&=&2k(\langle\Psi^l_1|W|\Psi^l_1\rangle+\langle\Psi^l_2|W|\Psi^l_2\rangle).
\end{eqnarray}

The equations for off-diagonal elements of $W$ with nonzero right-hand sides are
\begin{equation}
\langle\Psi^{\pm}_{n_1n_2}|\dot W(t)|\Psi^{\mp}_{n_1n_2}\rangle = -2k(n_1+n_2+1)\langle\Psi^{\pm}_{n_1n_2}|W|\Psi^{\mp}_{n_1n_2}\rangle.\nonumber
\end{equation}
The solutions of equations  (8) are
$$\langle\Psi^{\pm}_{n_1n_2}|W(t)|\Psi^{\mp}_{n_1n_2}\rangle =
\langle\Psi^{\pm}_{n_1n_2}|W(0)|\Psi^{\mp}_{n_1n_2}\rangle \exp\left\{- 2kt(n_1+n_2+1)\right\}$$ and the solutions of equations (5)-(7) may be obtained only numerically.
 For this purpose one can assumed that initially there is an
upper limit on the number of photons $N_1$ and $N_2$ in both of cavity modes so that
 $ \langle\Psi^{\pm}_{n_1 n_2}|W(t)|\Psi^{\mp}_{n_1 n_2}\rangle = 0 $
 for $n_1 >N_1, n_2 > N_2$. This implies
 that these matrix elements are zero for all $t$ since the cavity cannot add to the photon numbers. Then, one
 can start with $n_1=N_1+1$ and $n_2 = N_2 +1$ and iterate equations (5), (6) for smaller values of photon numbers untill $n_1=n_2=0$.
If there is no upper limit on the initial numbers of photons in the system the numbers $N_1$ and $N_2$
 must be taken large enough for the mean values of observables to calculate with the
 appropriate accuracy.
 These quantities may be obtained in the standard manner
$$\langle O(t)\rangle  = Sp \,O(t) \,W(t). $$

The solutions of the Eqs. (5)-(7) for arbitrary initial states of
atom and field can result from numerical calculations.

We consider below the NTPJCM with the atom initially in the excite state and the fields in coherent  states.

\section{Results and discussions for coherent input }

 The initial density matrix $W(0)$ for atom in the excited state and the
  fields in the coherent states is
\begin{eqnarray}
W(0)&=&\frac{p_{n_1}p_{n_2}}{2}\left(\gamma_{n_1n_2}^{+^2}|\Psi_{n_1n_2}^+\rangle
\langle\Psi_{n_1n_2}^+|+\gamma_{n_1n_2}^{-^2}|\Psi_{n_1n_2}^-\rangle
\langle\Psi_{n_1n_2}^-|+\right.\nonumber\\
&+&\left.\gamma_{n_1n_2}^+\gamma_{n_1n_2}^-|\Psi_{n_1n_2}^+\rangle \langle\Psi_{n_1n_2}^-|+\gamma_{n_1n_2}^-\gamma_{n_1n_2}^+|\Psi_{n_1n_2}^-\rangle \langle\Psi_{n_1n_2}^+|\right),\nonumber
\end{eqnarray}
where
$$p_{n_i} = exp(- n_i )\frac{\bar n_i^{n_i}}{n_i!}\quad (i=1,2).$$

First consider the time behaviour of   mean photon numbers and mean atomic populations
\begin{eqnarray}
\langle N_i(t)\rangle&=&\sum_{n_1=0}^{\infty}\sum_{n_2=0}^{\infty} \left\{\vphantom{\frac12} -\gamma_{n_1n_2}^+\gamma_{n_1n_2}^- \langle\Psi_{n_1n_2}^+|W|\Psi_{n_1n_2}^-\rangle
cos(2\Omega_{n_1n_2}t)+\right.\nonumber\\
 &+&\left(
n_i+\frac{\gamma_{n_1n_2}^{+^2}}{2}\right)
\langle\Psi_{n_1n_2}^-|W|\Psi_{n_1n_2}^-\rangle
+\left.\left(n_i+\frac{\gamma_{n_1n_2}^{-^2}}{2}\right)
\langle\Psi_{n_1n_2}^+|W|\Psi_{n_1n_2}^+\rangle
\right\}+\nonumber\\
&+&\langle\Psi_i^l|W|\Psi_i^l\rangle \quad (i=1,2),\nonumber\\
\langle
R_{e}(t)\rangle&=&\sum_{n_1=0}^{\infty}\sum_{n_2=0}^{\infty}
\left\{\frac{\gamma_{n_1n_2}^{+^2}}{2}
\langle\Psi_{n_1n_2}^+|W|\Psi_{n_1n_2}^+\rangle+
\frac{\gamma_{n_1n_2}^{-^2}}{2}
\langle\Psi_{n_1n_2}^-|W|\Psi_{n_1n_2}^-\rangle+\right.\nonumber\\
&+&\gamma_{n_1n_2}^+\gamma_{n_1n_2}^-\langle\Psi_{n_1n_2}^+|W|\Psi_{n_1n_2}^-\rangle
cos(2\Omega_{n_1n_2}t)\left.\vphantom{\frac12}\right\},\nonumber\\
\langle
R_{g}(t)\rangle&=&\sum_{n_1=0}^{\infty}\sum_{n_2=0}^{\infty}
\left\{\frac{\gamma_{n_1n_2}^{-^2}}{2}
\langle\Psi_{n_1n_2}^+|W|\Psi_{n_1n_2}^+\rangle+
\frac{\gamma_{n_1n_2}^{+^2}}{2}
\langle\Psi_{n_1n_2}^-|W|\Psi_{n_1n_2}^-\rangle-\right.\nonumber\\
&-&\gamma_{n_1n_2}^+\gamma_{n_1n_2}^-\langle\Psi_{n_1n_2}^+|W|\Psi_{n_1n_2}^-\rangle cos(2\Omega_{n_1n_2}t)\left.\vphantom{\frac12}\right\}+\sum_{z=1}^{3} \langle\Psi^l_z|W|\Psi^l_z\rangle.\nonumber
\end{eqnarray}\\

The mean populations of the excited atomic state  in the presence of the two modes of coherent state
 are plotted in Figs. 1 - 4 for  various values of  $\langle N_1\rangle, $ $\langle N_2\rangle$, $\delta$ and $k$.
 For small
 values of $\delta$ and $k$  the phenomena of quantum revivals and quantum collapses of the Rabi oscillation appear.
They are not as regular as those in one-photon or degenerate two-photon case.
  It can be seen that the amplitudes of the revival oscillations decrease as a
consequence of the cavity damping and detuning. In the case of strong damping, the cavity losses are so large that no collapses or revivals phenomena may appear. As a result of the computer simulations it can be said that the decay time and serene duration time for atomic populations
 are directly affected by the initial photon  numbers in the cavity modes.
 The serene duration time decreases as  $\langle N_1\rangle, $ $\langle N_2\rangle$ decreases and mean populations
  manifests the more fluctuating behaviour. For large field intensities
  the detuning influences both revival amplitudes and serene
duration time, as well as quasi-stationary atomic population value.

Since two photons are absorbed and/or emitted by the atom simultaneously in the cavity, one can tell that the behavior
 of photon numbers of mode 2 shows the exactly same manner as for the mode 1. Therefore in Figs.5 -8 we have plotted
 the mean photon number for first cavity mode. The mean photon numbers exhibit the same pattern of
 collapse and revival as the atomic population. A comparison of Figs. 1-8  shows that the photon numbers more significantly
 affected by cavity damping than the atomic population. In the case of strong damping or detuning the mean photon number
  decays exponentially.

  Perhaps a better appreciation of the statistics can be had by examining second-order
  correlation  function $G^{(2)}$ as a function
  of time t.
  The second-order correlation  functions for two cavity fields may be defined as
  $$
  G_i^{(2)}(t) = \frac{\langle(a^+(t))^2 a^2(t)\rangle - \langle a^+(t) a(t)\rangle^2}{\langle a^+(t) a(t)\rangle^2}.
  $$
  For strictly coherent field $G^{(2)}(0)=0$ whereas negative values of $G^{(2)}$
  lead to the antibunching of the field.
In dressed-state representation the second-order correlation  function becomes
  \begin{eqnarray}
G^{(2)}(t)&=&\frac{1}{\langle N_i(t)\rangle^2}\sum_{n_1=0}^{\infty} \sum_{n_2=0}^{\infty}\left\{ \left(n_1^2-n_1\frac{\Delta}{2\Omega_{n_1n_2}}\right)
\langle\Psi_{n_1n_2}^+|W|\Psi_{n_1n_2}^+\rangle\right.\nonumber\\
&+&\left(n_1^2+n_1\frac{\Delta}{2\Omega_{n_1n_2}}\right)
\langle\Psi_{n_1n_2}^+|W|\Psi_{n_1n_2}^+\rangle-\nonumber\\
&-&2n_i\gamma_{n_1n_2}^+\gamma_{n_1n_2}^- \langle\Psi_{n_1n_2}^+|W|\Psi_{n_1n_2}^-\rangle cos(2\Omega_{n_1n_2}t)\left.\vphantom{\frac12}\right\}-1.\nonumber
\end{eqnarray}

In Figs. 9 and 10 we plot $G_1^{(2)}$ for different detuning and damping parameters and large initial field intensities. For undamped resonant cavity collapses and revivals appear.
 In the case of  collapse in the absence of detuning (or cavity damping),
  $G_1^{(2)}\lesssim 0$ , the  oscillations show both bunching and antibunching features.
 In the  cases of nonzero damping and detuning the antibunching effects almost disappear.
 The amplitudes of the revivals oscillations  will be damped by small cavity losses to the
 extent that antibunching appears only at the very beginning of the time
 evolution.  For large detuning the second-order correlation function $G_1^{(2)} > 0$
  for every $t$.

Finally, we study the field and atomic squeezing. In order to investigate the squeezing properties of the radiation field
 we define the slowly varying Hermitian quadrature operators for fields
$$X^{(i)}_1=\frac{1}{2}(a_i e^{\imath\omega_i t} + a_i^+e^{-\imath \omega_i t}),$$
$$X^{(i)}_2=\frac{1}{2\imath}(a_i e^{\imath\omega_i t} +-a_i^+e^{-\imath \omega_i t})$$
$$ \quad (i=1,2).$$
The commutation of $X^{(i)}_1$ and $X^{(i)}_2$ is $[X^{(i)}_1, X^{(i)}_2]=\imath/2$. The variances $(\Delta X^{(i)}_j)^2 = \langle (X^{(i)}_j)^2 \rangle - \langle X^{(i)}_j \rangle^2 \quad (j=1,2)$ satisfy
 the uncertainty relation $(\Delta X^{(i)}_1)^2 (\Delta X^{(i)}_2)^2 \geq 1/16$. For the vacuum and coherent states of
 the field the variances are equal 1/4. The field is in a squeezed state if there takes
 place $(\Delta X^{(i)}_j)^2 < 1/4$ for either $j=1$ or $2$.

 The condition for squeezing in the $j$th quadrature $\Delta X^{(i)}_j$ can be written simply as
 $$S^{(i)}_j < 1, $$
 where the squeezing factor is
 $$S^{(i)}_j = 4 \Delta X^{(i)}_j  \qquad (j=1,2).$$
For the sake of definiteness we study the  squeezing properties of  the first  cavity mode.
 In terms of the photon operators, one can readily find  that the squeezing parameter
  of the first quadrature component and for the first cavity mode may be
  written as
 $$S = \langle a^2_1\rangle+\langle a^{+^2}_1\rangle+2\langle a^+_1a_1\rangle-
 (\langle a^+_1+a_1\rangle)^2+1,$$
 where
\begin{eqnarray}
 \langle a^2_1\rangle+\langle a^{+^2}_1\rangle &=&
 \sum_{n_1,n_2=0}^{\infty}\sqrt{n_1+2}\left\{
 (\gamma_{n_1n_2}^+\gamma_{n_1+2\,n_2}^+\sqrt{n_1+1}+
\gamma_{n_1n_2}^-\gamma_{n_1+2\,n_2}^-\sqrt{n_1+3})\times
\right.\nonumber\\
&\times& \langle\Psi_{n_1+2\,n_2}^+|W|\Psi_{n_1n_2}^+\rangle
 cos([\Omega_{n_1+2\,n_2}-\Omega_{n_1n_2}]t)+\nonumber\\
&+&(\gamma_{n_1n_2}^+\gamma_{n_1+2\,n_2}^-\sqrt{n_1+1}- \gamma_{n_1n_2}^-\gamma_{n_1+2\,n_2}^+\sqrt{n_1+3})\times
\nonumber\\
&\times&\langle\Psi_{n_1+2\,n_2}^-|W|\Psi_{n_1n_2}^+\rangle
 cos([\Omega_{n_1+2\,n_2}+\Omega_{n_1n_2}]t)+\nonumber\\
&+& (\gamma_{n_1n_2}^-\gamma_{n_1+2\,n_2}^-\sqrt{n_1+1}+ \gamma_{n_1n_2}^+\gamma_{n_1+2\,n_2}^+\sqrt{n_1+3})\times
\nonumber\\
&\times&\langle\Psi_{n_1+2\,n_2}^-|W|\Psi_{n_1n_2}^-\rangle
 cos([\Omega_{n_1+2\,n_2}-\Omega_{n_1n_2}]t)+\nonumber\\
&+& (\gamma_{n_1n_2}^-\gamma_{n_1+2\,n_2}^+\sqrt{n_1+1}-
\gamma_{n_1n_2}^+\gamma_{n_1+2\,n_2}^-\sqrt{n_1+3})\times\nonumber\\
&\times& \langle\Psi_{n_1+2\,n_2}^+|W|\Psi_{n_1n_2}^-\rangle
 cos([\Omega_{n_1+2\,n_2}+\Omega_{n_1n_2}]t)\left.\vphantom{gt}\right\},\nonumber
\end{eqnarray}
\begin{eqnarray}
 \langle a_1\rangle+\langle a^{+}_1\rangle &=&
 \sum_{n_1,n_2=0}^{\infty}\left\{
 (\gamma_{n_1n_2}^+\gamma_{n_1+1\,n_2}^+\sqrt{n_1+1}+
\gamma_{n_1n_2}^-\gamma_{n_1+1\,n_2}^-\sqrt{n_1+2})\times
\right.\nonumber\\
&\times& \langle\Psi_{n_1+1\,n_2}^+|W|\Psi_{n_1n_2}^+\rangle
 cos([\Omega_{n_1+1\,n_2}-\Omega_{n_1n_2}]t)+\nonumber\\
&+& (\gamma_{n_1n_2}^+\gamma_{n_1+1\,n_2}^-\sqrt{n_1+1}- \gamma_{n_1n_2}^-\gamma_{n_1+1\,n_2}^+\sqrt{n_1+2})\times
\nonumber\\
&\times&\langle\Psi_{n_1+1\,n_2}^-|W|\Psi_{n_1n_2}^+\rangle
 cos([\Omega_{n_1+1\,n_2}+\Omega_{n_1n_2}]t)+\nonumber\\
&+& (\gamma_{n_1n_2}^-\gamma_{n_1+1\,n_2}^-\sqrt{n_1+1}+ \gamma_{n_1n_2}^+\gamma_{n_1+1\,n_2}^+\sqrt{n_1+2})\times
\nonumber\\
&\times&\langle\Psi_{n_1+1\,n_2}^-|W|\Psi_{n_1n_2}^-\rangle
 cos([\Omega_{n_1+1\,n_2}-\Omega_{n_1n_2}]t)+\nonumber\\
&+& (\gamma_{n_1n_2}^-\gamma_{n_1+1\,n_2}^+\sqrt{n_1+1}-
\gamma_{n_1n_2}^+\gamma_{n_1+1\,n_2}^-\sqrt{n_1+3})\times\nonumber\\
&\times& \langle\Psi_{n_1+1\,n_2}^+|W|\Psi_{n_1n_2}^-\rangle
 cos([\Omega_{n_1+1\,n_2}+\Omega_{n_1n_2}]t)\left.\vphantom{gt}\right\},\nonumber
\end{eqnarray}
and $\langle a_1^+a_1\rangle=\langle N_1(t)\rangle$.

 Values $S(t)<1$ imply squeezing in the first
quadrature component.  In Figs. 11 and 12  we have plotted $S(t)$ for different  values of $k$ and $\delta$ and large input field intensities.
 For $k=0$ there is some amount of squeezing which appears only at the very beginning
 of the time and this decreases with increasing of $\delta$ in contrast with degenerate two-photon JCM.
 But this model does not give rise to significant
 squeezing.  For nonideal cavity the effect is vanished. So,
 the cavity damping is seen to have an appreciable effect on the squeezing properties.

 To investigate atomic squeezing we introduce the dispersive and absorptive  components
  of slowly varying atomic dipole moment \cite{32}
 $$\sigma_1 =\frac{1}{2}(R^+ e^{-\imath \omega_0 t} + R^- e^{\imath \omega_0 t})$$
 and
$$\sigma_2 =\frac{1}{2\imath}(R^+ e^{-\imath \omega_0 t} - R^- e^{\imath \omega_0 t}),$$
respectively. They obey the commutation relation
$$[\sigma_1, \sigma_2] =\frac{1}{2} \sigma_3$$
and the corresponding uncertainty relation
$$(\Delta \sigma_1)^2 (\Delta \sigma_2)^2 \geq \frac{1}{16}\langle \sigma _3\rangle^2.$$
The atomic state is said to be squeezed when $\sigma_1$ or $\sigma_2$ satisfies the relation

$$(\Delta \sigma_i)^2 < \frac{1}{4} |\langle \sigma_3 \rangle |, \quad (i=1,2)\eqno{(9)}$$
Since
$$
(\Delta \sigma_1)^2 =\frac{1}{4}-(Re \langle \sigma\rangle e^{\imath \omega_0 t})^2, $$
$$(\Delta \sigma_2)^2 =\frac{1}{4}-(Im \langle \sigma\rangle e^{\imath \omega_0 t})^2,$$
the condition described by Eq. (9) can be rewritten as
$$
F_1 = \frac{1 - 4 (Re \langle \sigma\rangle e^{-\imath \omega_0 t})^2}{|\langle \sigma_3\rangle|} < 1$$ or
$$
F_2 = \frac{1 - 4 (Im \langle \sigma\rangle e^{-\imath \omega_0 t})^2}{|\langle \sigma_3\rangle|} < 1.$$ for squeezing in the dispersive or absorptive component of the dipole moment.

Here
\begin{eqnarray}
\langle \sigma \rangle e^{-\imath \omega_0 t} &=& \frac{e^{\imath \Delta t}}{2} \sum_{n_1,n_2=0}^{\infty}\left\{\gamma_{n_1+1\,n_2+1}^{+}\gamma_{n_1\,n_2}^{-} e^{-\imath t\left(\Omega_{n_1+1\,n_2+1}-\Omega_{n_1\,n_2}\right)} \langle \Psi_{n_1\,n_2}^{+}|W|\Psi_{n_1+1\,
n_2+1}^{+}\rangle \right.-\nonumber \\
&-&\gamma_{n_1+1\,n_2+1}^{+}\gamma_{n_1\,n_2}^{+} e^{-\imath t\left(\Omega_{n_1+1\,n_2+1}+\Omega_{n_1\,n_2}\right)} \langle \Psi_{n_1\,n_2}^{-}|W|\Psi_{n_1+1\,
n_2+1}^{+}\rangle +\nonumber \\
&+&\gamma_{n_1+1\,n_2+1}^{-}\gamma_{n_1\,n_2}^{-} e^{\imath t\left(\Omega_{n_1+1\,n_2+1}+\Omega_{n_1\,n_2}\right)} \langle \Psi_{n_1\,n_2}^{+}|W|\Psi_{n_1+1\,
n_2+1}^{-}\rangle-\nonumber \\
&-&\gamma_{n_1+1\,n_2+1}^{-}\gamma_{n_1\,n_2}^{+} e^{\imath t\left(\Omega_{n_1+1\,n_2+1}-\Omega_{n_1\,n_2}\right)} \langle \Psi_{n_1\,n_2}^{-}|W|\Psi_{n_1+1\, n_2+1}^{-}\rangle \left.\vphantom{\frac12}\right\},\nonumber
\end{eqnarray}
and $\langle \sigma_3 \rangle = \langle R_e\rangle-\langle R_g \rangle$ is atomic inversion in terms of dressed states representation.

 The results for the squeezing parameters for different values
of $k$ and $\delta$ and moderate input intensities ($\langle n_1\rangle = 15, \langle n_2\rangle = 10)$ have been shown on Figs. 13-16. The dispersive component $F_1$ does not squeeze at the very beginning of the time, the absorptive component $F_2$, on the other hand, goes below 1 with virtually no time delay.

 Both $F_1$ and $F_2$ shows squeezing recurrently  only for small times
 of the atom-field  interaction. As that has been mentioned earlier \cite{23} that
  squeezing does not show up in either case until after
 $\langle N_i \rangle \geq 7.0 $. The amount of squeezing decreases with increase
  of damping parameter $k$. If the parameter $k$ is  large enough
  the squeezing in dispersive component vanishes and the squeezing in absorptive component
   of dipole moment  undergoing only one squeezing minimum. For
 high input intensities (not shown in Graphs) the influence of damping parameter
  on squeezing amount
is more dramatic than one for medium input intensities. And with increasing the detuning parameter $\delta$ the amount of squeezing in dispersive and absorptive component also decreases while the time interval, for which the squeezing appears,
 increases.

\section{Summary}
In this paper we have investigated the  nondegenerate two-photon Jaynes-Cummings model  with damping and detuning. We have set kinetic equations for density matrix of the considered system in secular approximation and with using the dressed-state representation. These equations are solved numerically for different detuning of atomic levels and damping parameter. On the basis of these equation the analysis of the dynamical
 behaviour of mean values of atomic populations, mean photon numbers, field coherence and squeezing
 has been carried out.
The effects of cavity damping will significantly attenuate the amplitudes of mean atomic populations,
 mean photon number revivals. The revivals of the second-order photon correlation function will be damped by
 small cavity losses to the extent than antibunching appears only at the very beginning of the time. Cavity damping
 has an appreciable effect on
 the squeezing properties of fields and atomic dipole moment. For moderate damping parameter the field squeezing
  is vanished and the amount of atomic squeezing sharply decreases. In this paper we shall restrict our consideration to
  the coherent input for  fields and  ignore  the Stark shift.
   A further discussion on dissipative NTPJCM including the consideration of
  initial squeezed and thermal states for cavity fields and Stark shift is planned to be reported in the subsequent
  paper.

\section{Appendix}

\begin{figure}[!h]
\resizebox{115mm}{65 mm}{\includegraphics{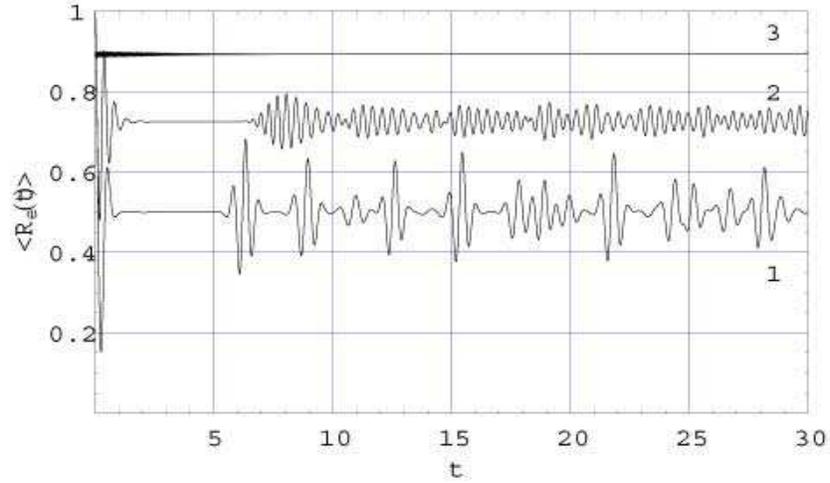}}
 \caption{Atomic population of an excited level for
   $\langle N_1\rangle=\langle N_2\rangle=5,\,  \delta=10$ and {\bf 1.}
   $k=0$, {\bf 2.} $k=0.001$, {\bf 3.} $k=0.01$. \,
    Curve 3 corresponds to value  $\langle R_e(t)\rangle-0.1$}
\end{figure}
\begin{figure}[!h]
\resizebox{115mm}{65 mm}{\includegraphics{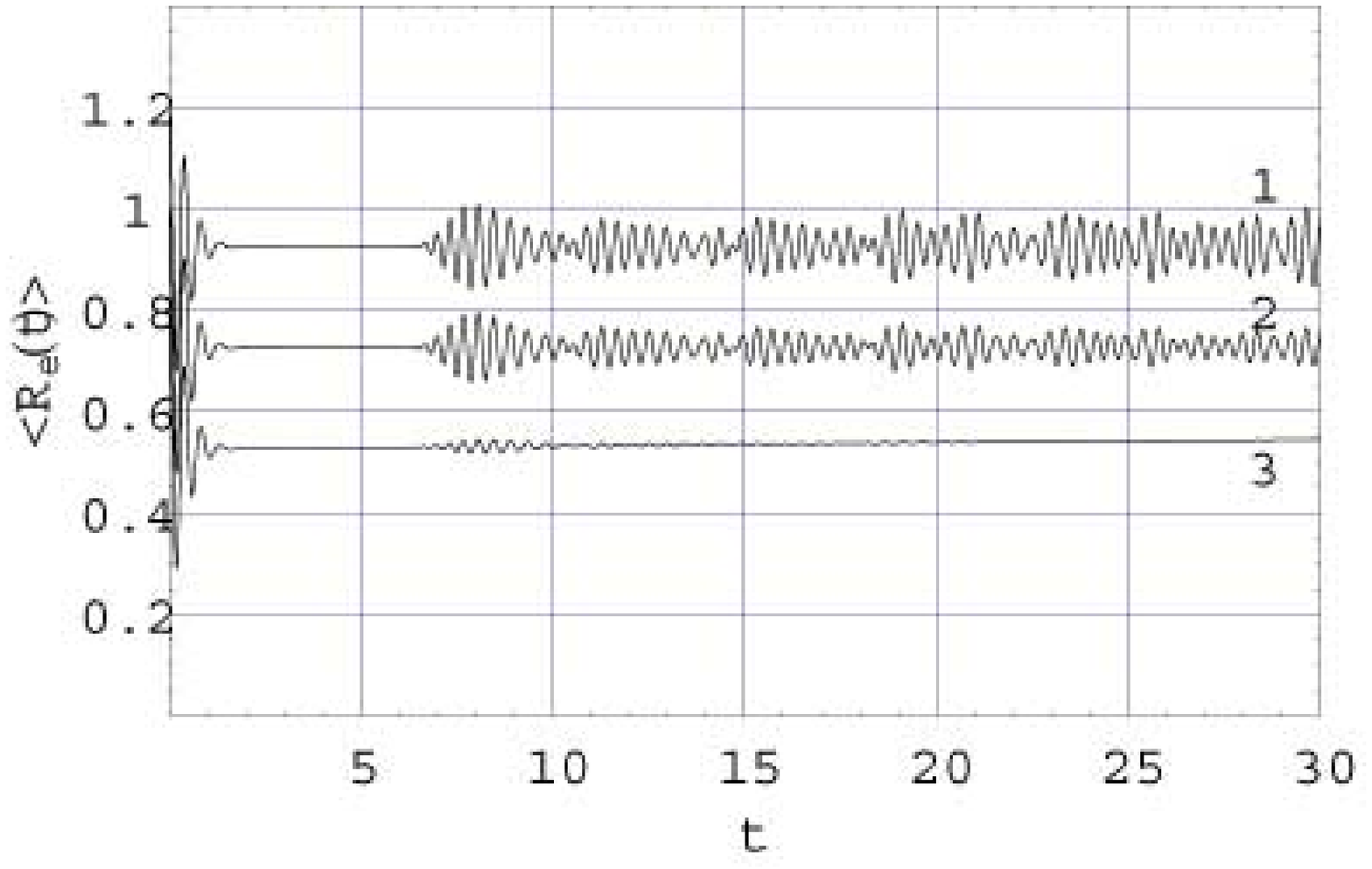}} \caption{Atomic population of an excited level
for $\langle N_1\rangle=\langle N_2\rangle=5,\, \delta=10$ and
{\bf 1.} $k=0$, {\bf 2.} $k=0.001$, {\bf 3.} $k=0.01$. Curve 1
corresponds to value $\langle R_e(t)\rangle+0.2$, and curve 3
corresponds to $\langle R_e(t)\rangle-0.2$}
\end{figure}

\begin{figure}[!h]
\resizebox{115mm}{65 mm}{\includegraphics{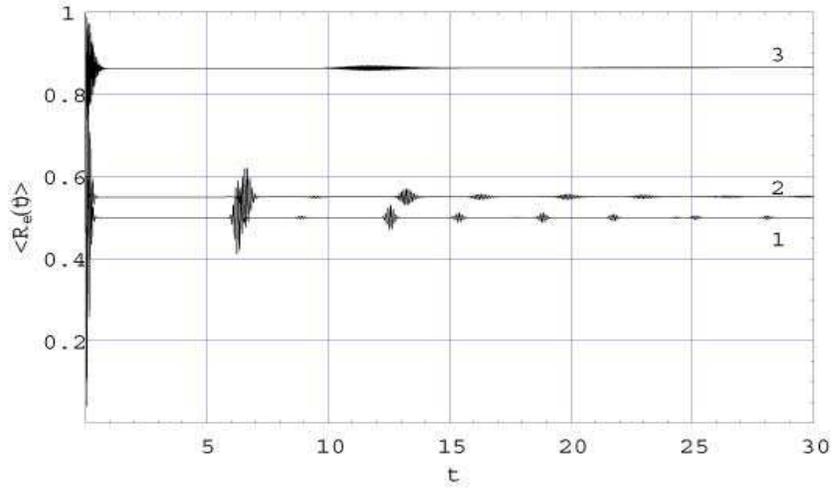}}\caption{Atomic population of an excited level
for  $\langle N_1\rangle=\langle N_2\rangle=30,\, k=0.001$ and
{\bf 1.} $\delta=0$; \,  {\bf 2.} $\delta=20$; {\bf 3.}
$\delta=100$.}
\end{figure}

\begin{figure}[!h]
\resizebox{115mm}{65 mm}{\includegraphics{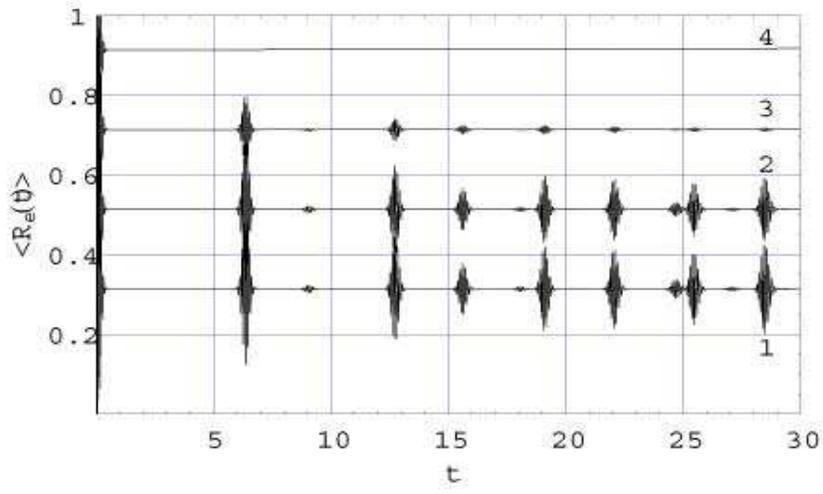}} \caption{Atomic population of an excited level
for $\langle N_1\rangle =\langle N_2\rangle =30,\, \delta=10$ and
{\bf 1.} $k=0$; \, {\bf 2.} $k=0.0001$;\, {\bf 3.}
$k=0.001$;\,{\bf 4.}\,$k=0.01$. Curve 1 corresponds to value
$\langle R_e(t)\rangle -0.2$,
 curve 3 corresponds to $\langle R_e(t)\rangle +0.2$, curve 4
 corresponds
 to $\langle\ R_e(t)\rangle +0.4$.}
\end{figure}
\begin{figure}[!h]
\resizebox{115mm}{65 mm}{\includegraphics{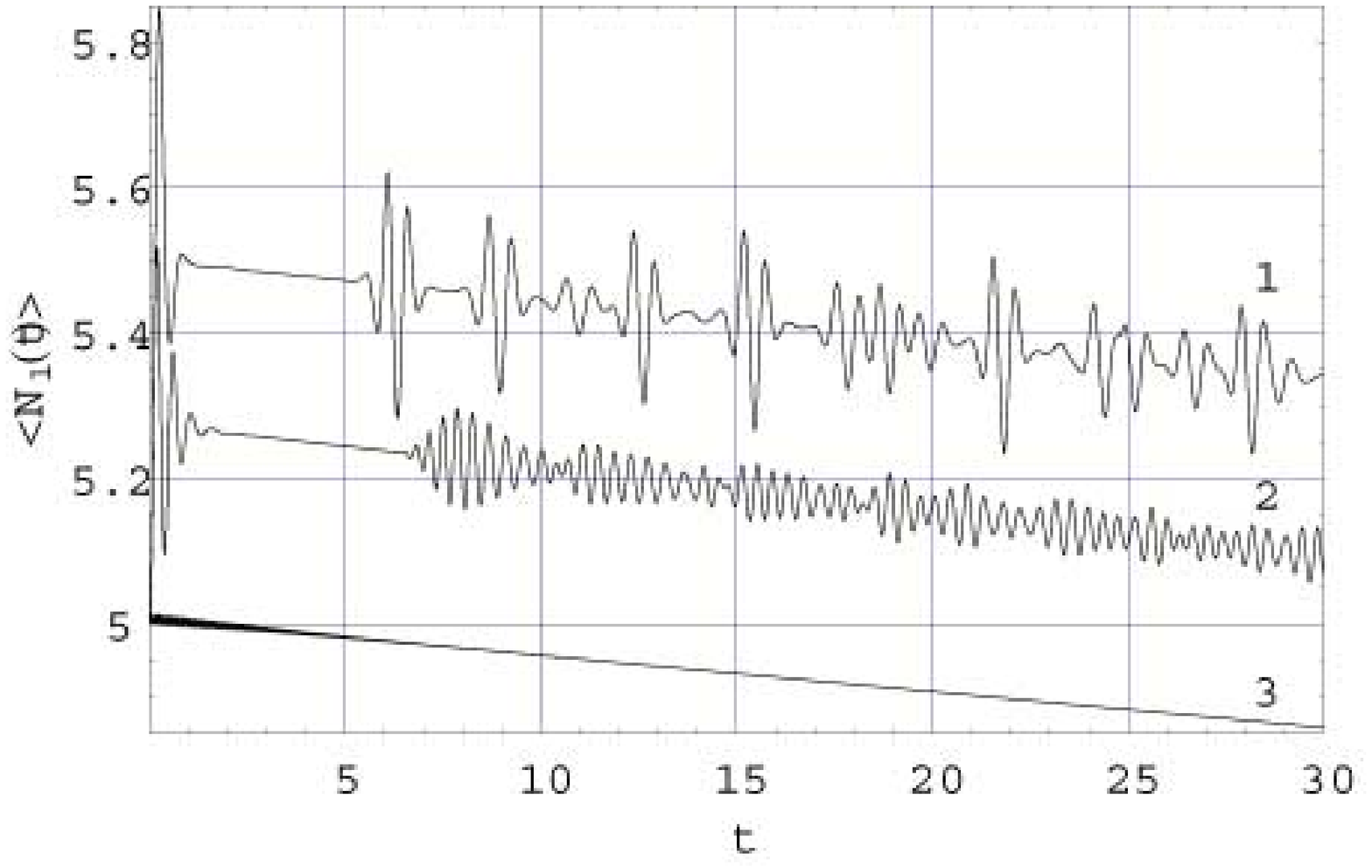}} \caption{Mean photon number in the first
field mode for $\langle N_1\rangle=\langle N_2\rangle=5,\,
k=0.001$ and {\bf 1.} $\delta=0$; {\bf 2.} $\delta=10$; {\bf 3.}
$\delta=100$.}
\end{figure}

\begin{figure}[!h]
\resizebox{115mm}{65 mm}{\includegraphics{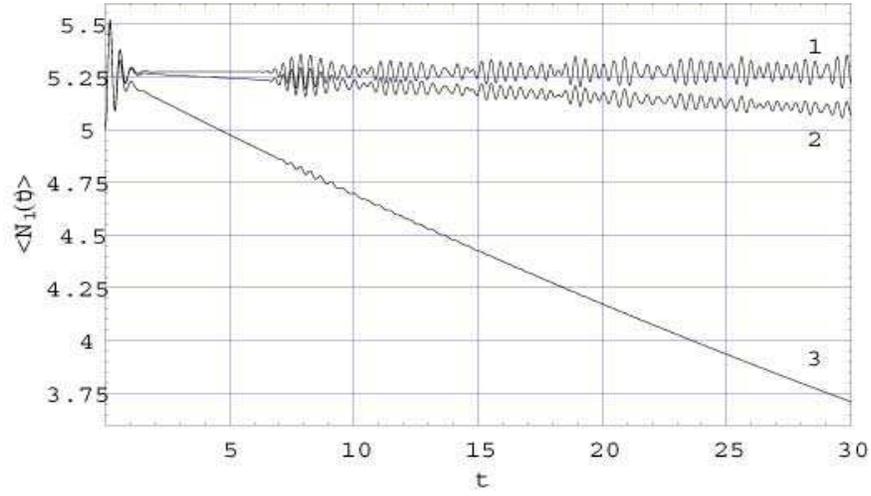}} \caption{Mean photon number in the first field
mode for $\langle N_1\rangle=\langle N_2\rangle=5,\,  \delta=10$
and {\bf 1.} $k=0$, {\bf 2.} $k=0.001$, {\bf 3.} $k=0.01$. }
\end{figure}

\begin{figure}[!h]
\resizebox{115mm}{65mm}{\includegraphics{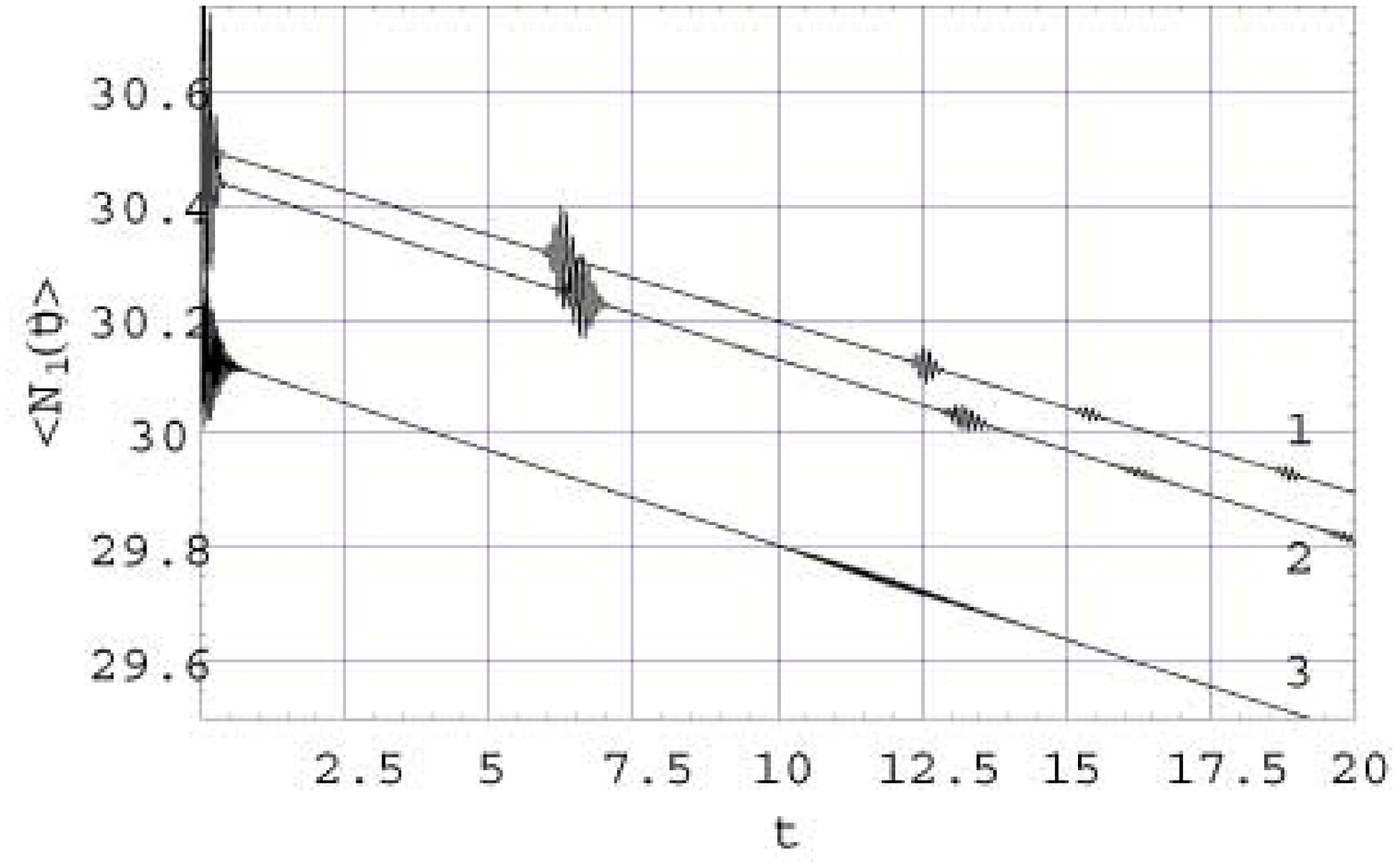}} \caption{Mean photon number in the first
field mode for  $\langle N_1\rangle=\langle N_2\rangle=30,\,
k=0.001$ and {\bf 1.} $\delta=0$; \,  {\bf 2.} $\delta=20$;\,
{\bf 3.} $\delta=100$.}
\end{figure}

\begin{figure}[!h]
\resizebox{115mm}{65 mm}{\includegraphics{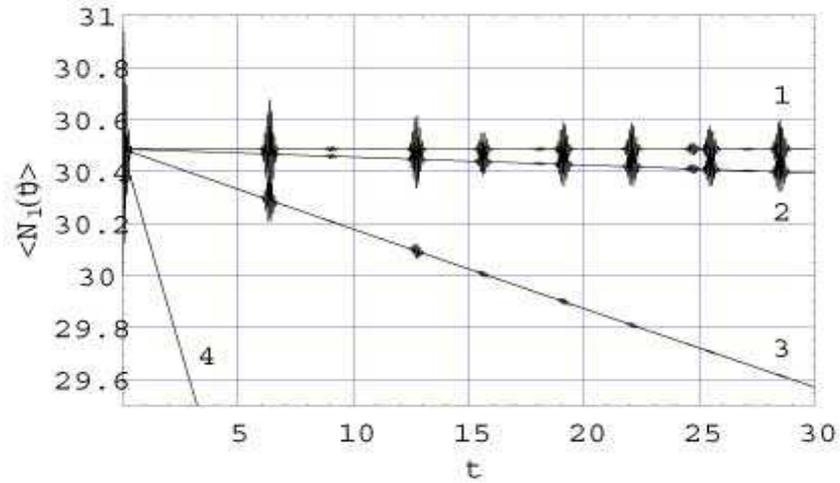}} \caption{Mean photon number in the first field
mode for  $\langle N_1\rangle=\langle N_2\rangle=30,\, \delta=10$
and {\bf 1.} $k=0$; \,  {\bf 2.} $k=0.0001$;\, {\bf 3.}
$k=0.001$;\,{\bf 4.}\,$k=0.01$.}
\end{figure}

\begin{figure}[!h]
\resizebox{115mm}{65 mm}{\includegraphics{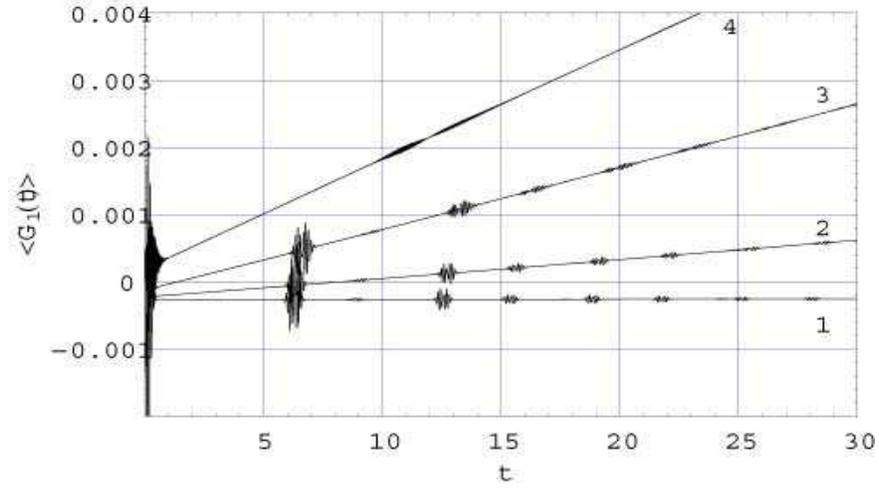}} \caption{Second order correlation function for
the first field mode for $\langle N_1\rangle=\langle
N_2\rangle=30,\, k=0.001$ and {\bf 1.} $\delta=0$; \,  {\bf 2.}
$\delta=10$;\, \,  {\bf 3.} $\delta=20$;\, {\bf 4.} $\delta=100$.}
\end{figure}

\begin{figure}[!h] \resizebox{115mm}{65
mm}{\includegraphics{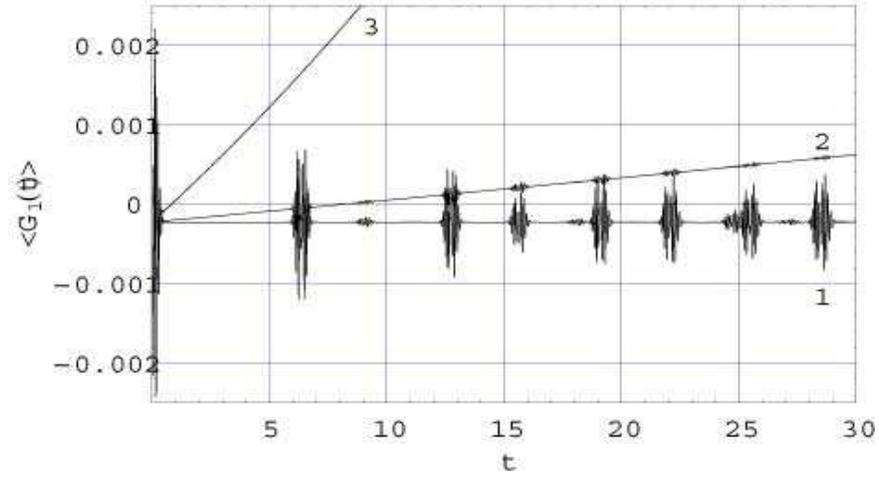}}
\caption{Second order correlation function for the first field
mode for $\langle N_1\rangle=\langle N_2\rangle=30,\, \delta=10$
and {\bf 1.} $k=0$; \,  {\bf 2.} $k=0.001$;\, {\bf 3.} $k=0.01$.}
\end{figure}

\begin{figure}[!h]
\resizebox{115mm}{65mm}{\includegraphics{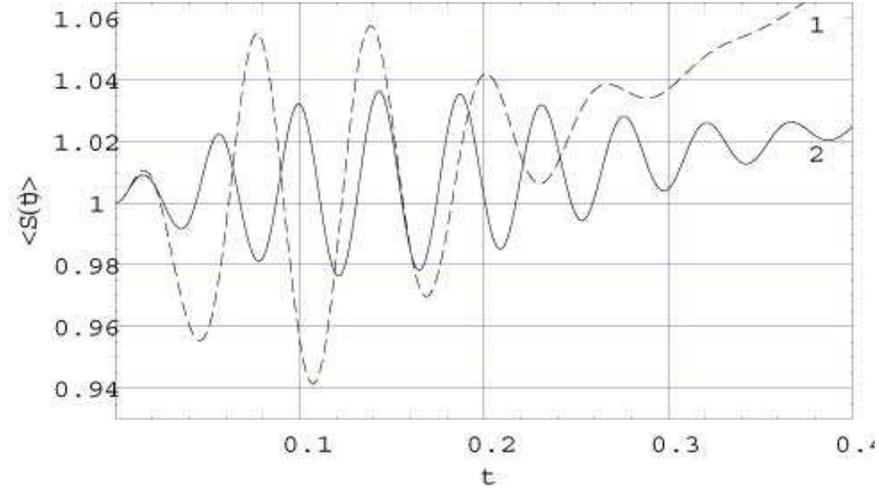}} \caption{Squeezing in the first field
mode for $\langle N_1\rangle=\langle N_2\rangle=50, k=0$ and {\bf
1.}\, $ \delta=10$; {\bf 2.}\, $ \delta=100$.}
\end{figure}

\begin{figure}[!h]
\resizebox{115mm}{65mm}{\includegraphics{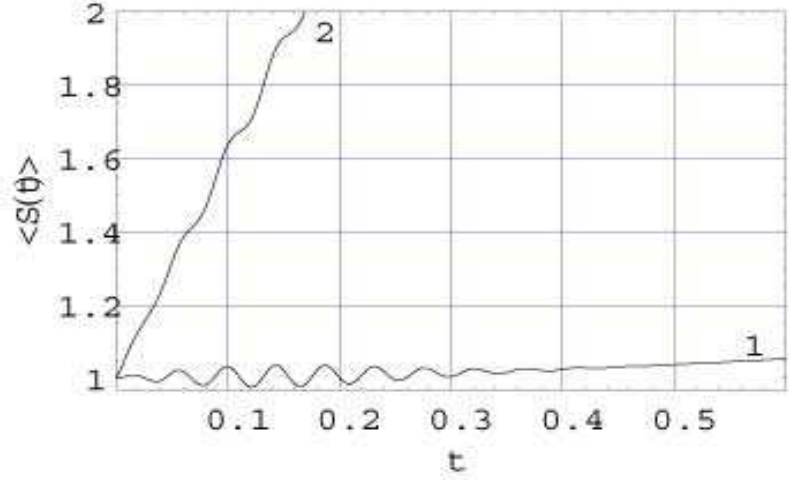}} \caption{Squeezing in the first field mode
for $\langle N_1\rangle=\langle N_2\rangle=50,\, \delta=10$ and
{\bf 1.} $k=0$; {\bf 2.}\,$ k=0.0001.$}
\end{figure}

\begin{figure}[!h]
\resizebox{115mm}{65mm}{\includegraphics{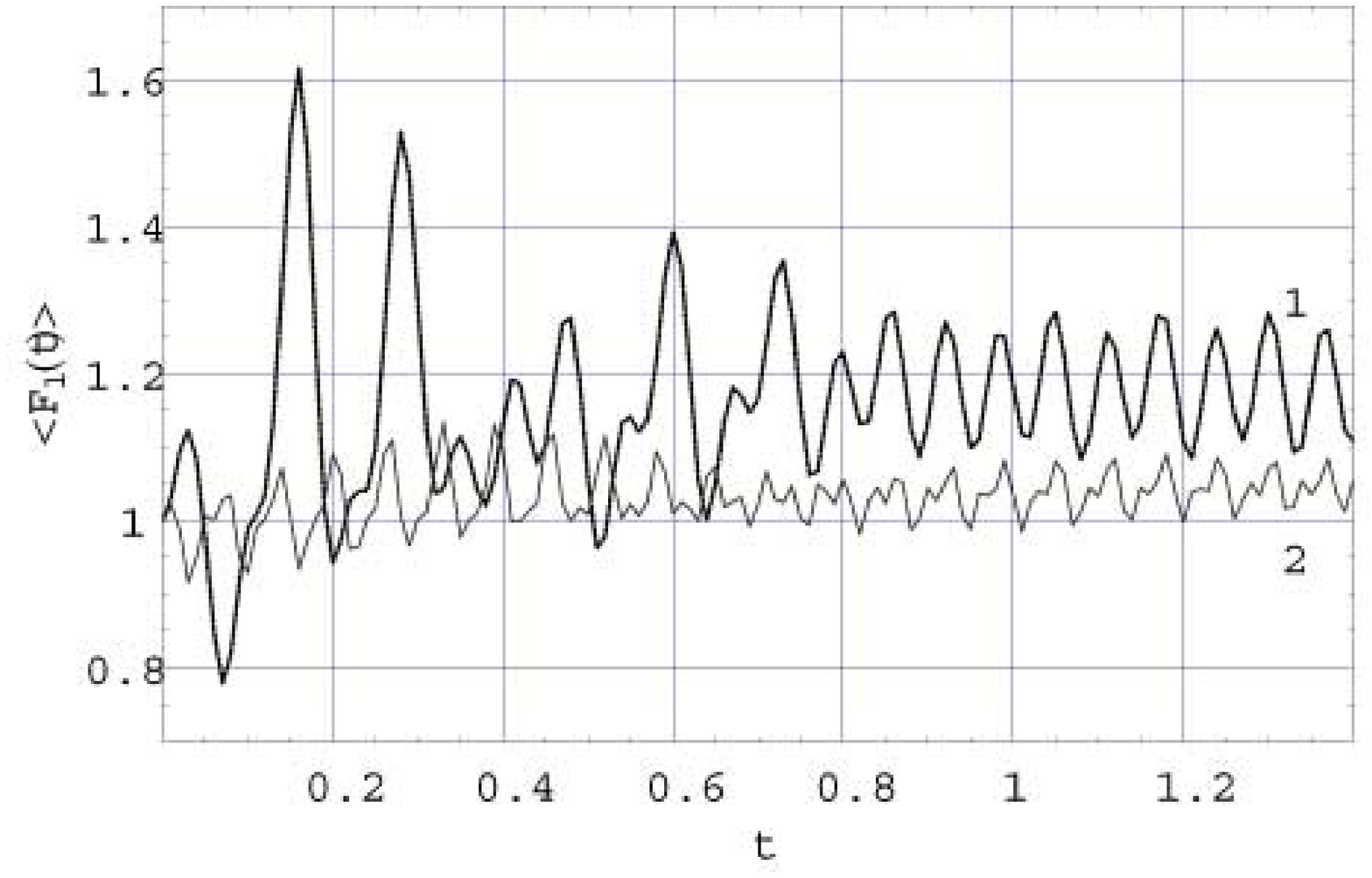}} \caption{Atomic dipole moment dispersive
component for $\langle N_1\rangle=15,\,$ $\langle N_2\rangle=10,$
\, $k=0.001$ and \,{\bf 1.}\, $\delta=50$ ; {\bf
2.}\,$\delta=100$.}
\end{figure}

\begin{figure}[!h]
\resizebox{115mm}{65mm}{\includegraphics{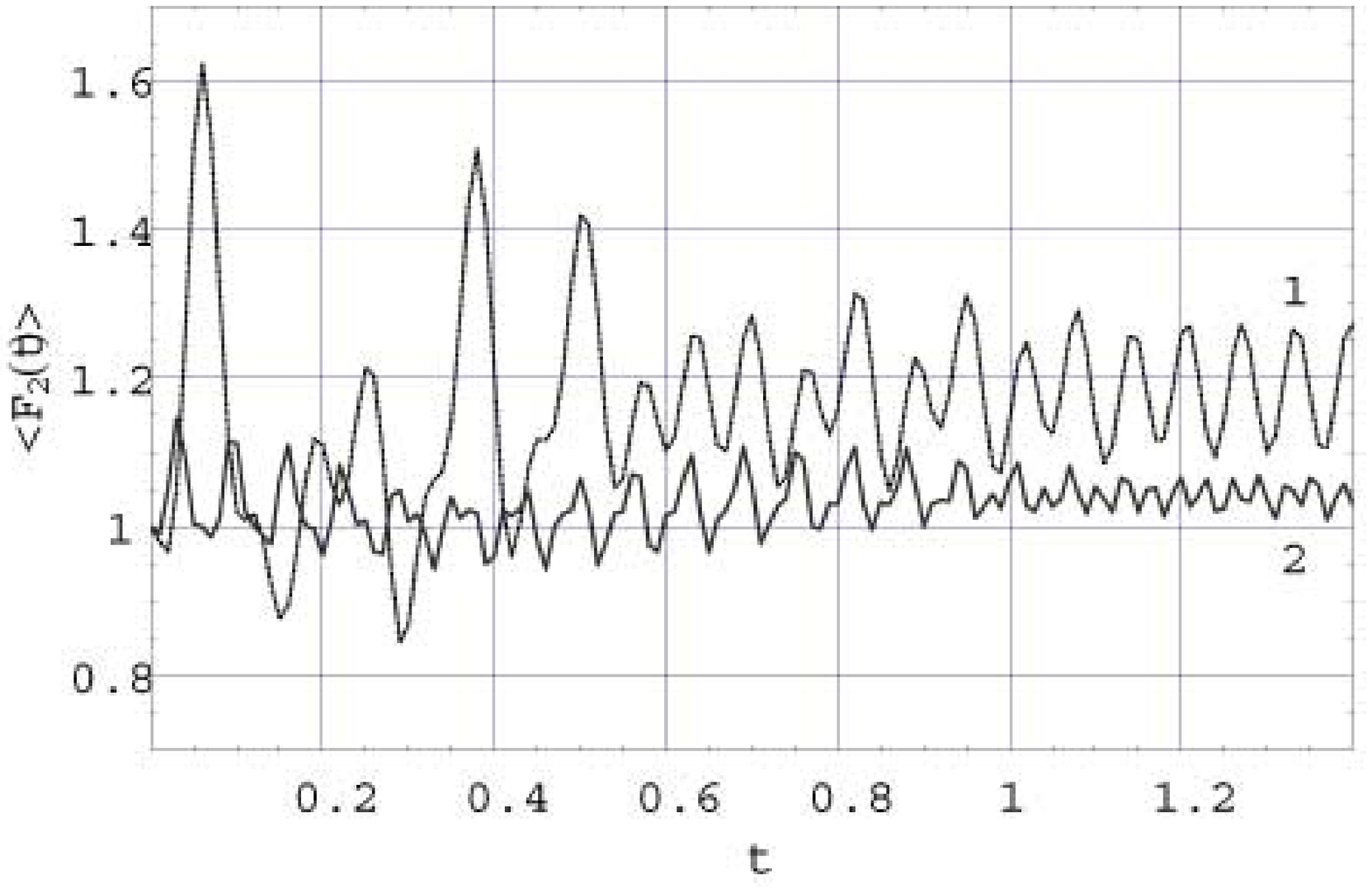}} \caption{Atomic dipole moment absorptive
component for $\langle N_1\rangle=15,\,$ $\langle N_2\rangle=10,$
\, $k=0.001$ and \,{\bf 1.}\, $\delta=50$;\, {\bf
2.}\,$\delta=100$.}
\end{figure}

\begin{figure}[!h]
\resizebox{115mm}{65mm}{\includegraphics{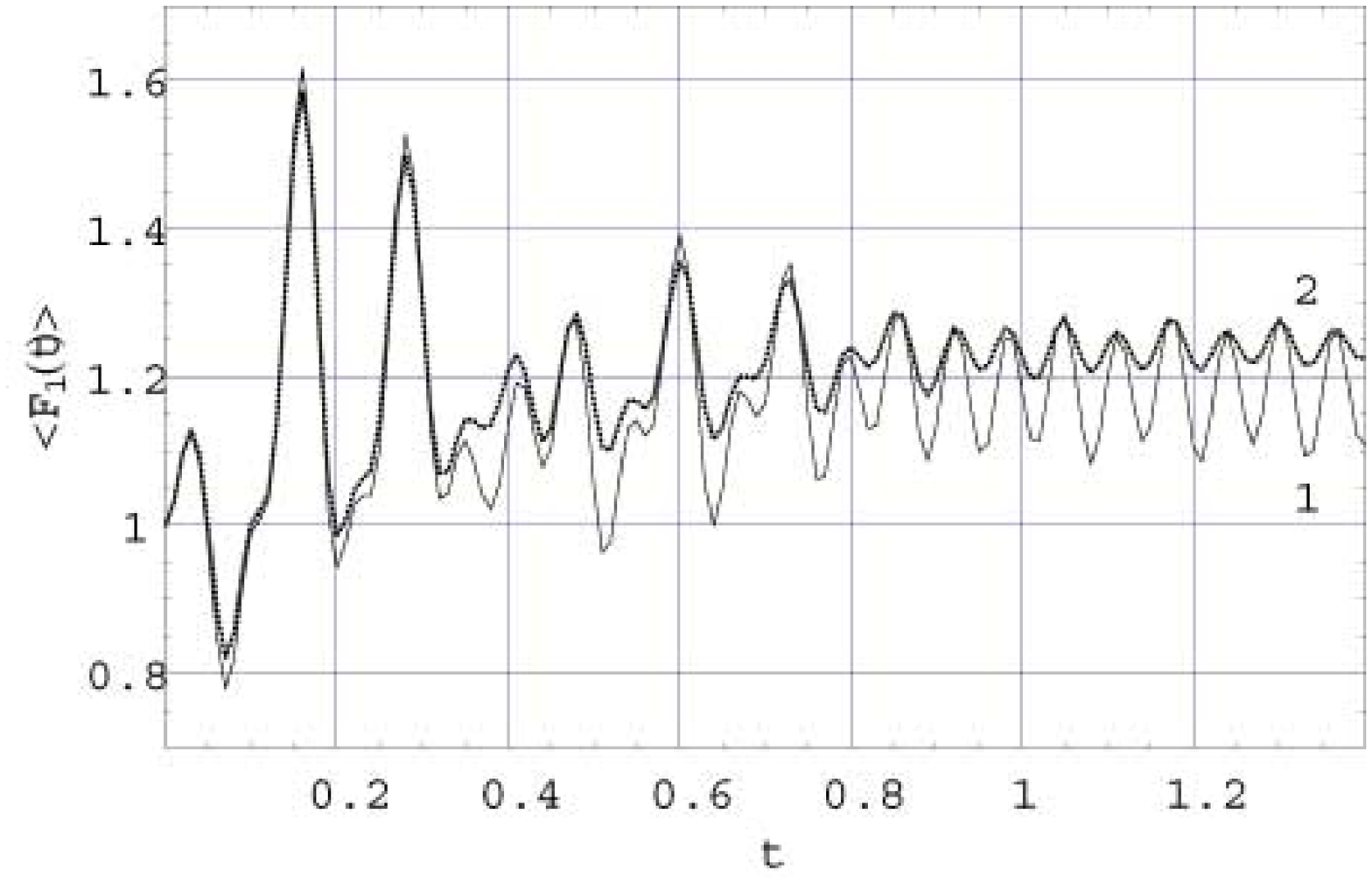}} \caption{Atomic dipole moment dispersive
component for $\langle N_1\rangle=15,\,$ $\langle N_2\rangle=10,$
\, $\delta=10$ and \,{\bf 1.}\, $k=0.001$;\, {\bf 2.}\,$k=0.01$.}
\end{figure}

\begin{figure}[t]
\resizebox{115mm}{65mm}{\includegraphics{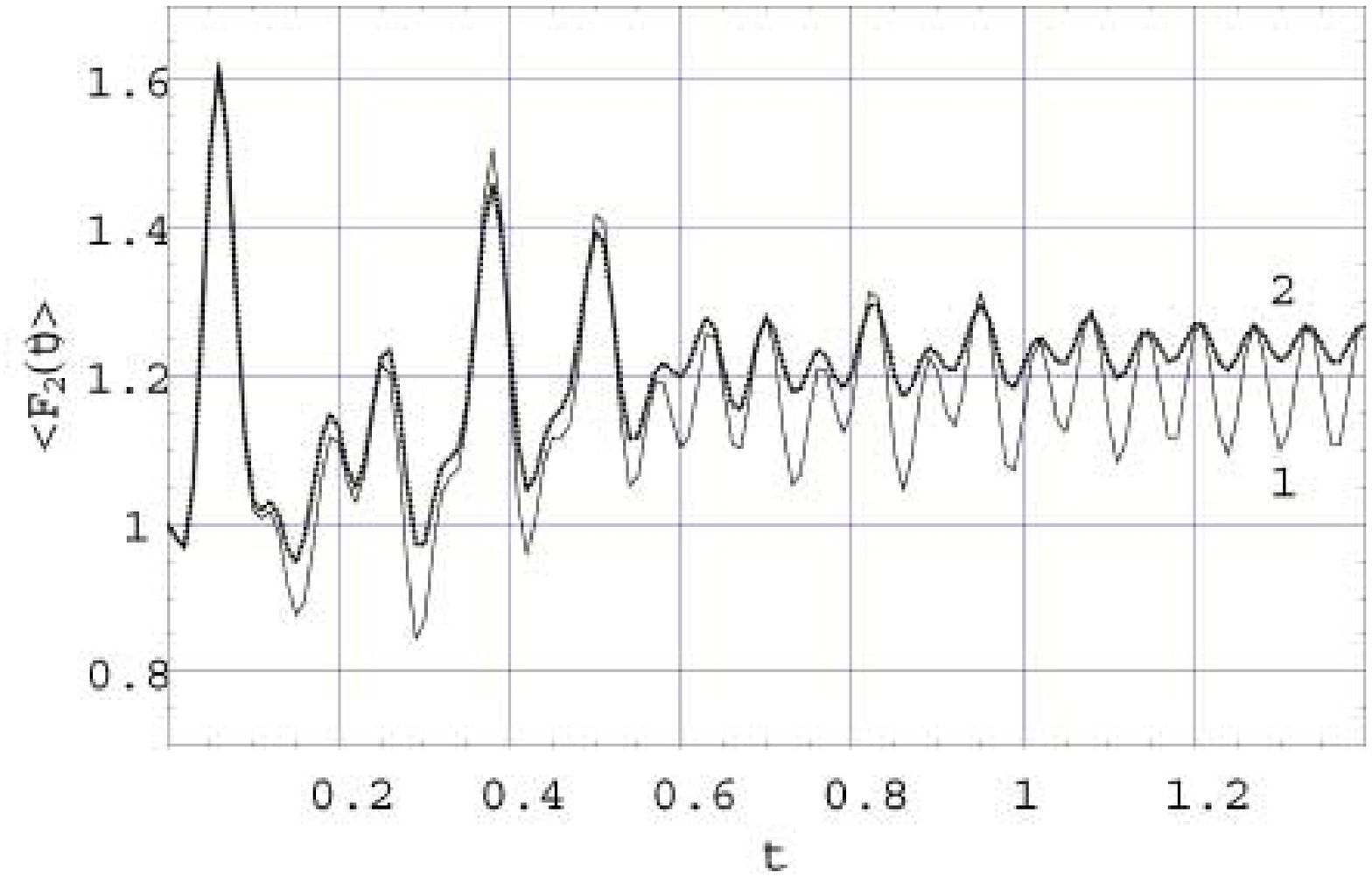}} \caption{Atomic dipole moment absorptive
component for $\langle N_1\rangle=15,\,$ $\langle N_2\rangle=10
,$\, $\delta=10$ and \,{\bf 1.}\, $k=0.001$;\, {\bf
2.}\,$k=0.01$.}
\end{figure}

\end{document}